\documentclass[aps,prl,preprint,tightenlines,superscriptaddress,showpacs,byrevtex]{revtex4}
\usepackage{graphicx} 
\usepackage{color}

\usepackage{dcolumn}  

\newcommand{\cm}{\mbox{~cm}}

\newcommand{\GeV}{\mbox{~GeV}}
\newcommand{\GeVc}{\GeV/c}
\newcommand{\GeVcc}{\GeVc^2}
\newcommand{\MeV}{\mbox{~MeV}}
\newcommand{\MeVc}{\MeV/c}
\newcommand{\MeVcc}{\MeVc^2}
\newcommand{\invfb}{\mbox{~fb}^{-1}}

\newcommand{\ks}{K^0_{S}}

\newcommand{\ee}{e^{+}e^{-}}

\newcommand{\etal}{\it et al.\rm}

\newcommand{\kpp}{K\pi^+\pi^-}
\newcommand{\kppp}{K^+\pi^+\pi^-}

\newcommand{\kspp}{K^0_{S}\pi^+\pi^-}
\newcommand{\kpppg}{\kppp\gamma}
\newcommand{\ksppg}{\kspp\gamma}
\newcommand{\mkpp}{M_{K\pi\pi} }

\newcommand{\konei}{K_1 (1270)}
\newcommand{\koneii}{K_1 (1400)}
\newcommand{\koneip}{K_1 (1270)^+}
\newcommand{\koneiip}{K_1 (1400)^+}
\newcommand{\ktwost}{K_2^* (1430)}
\newcommand{\ktwostp}{K_2^* (1430)^+}

\newcommand{\kppg}{K\pi^+\pi^-\gamma}

\newcommand{\bptokpppg}{B^+ \rightarrow K^+\pi^+\pi^-\gamma}
\newcommand{\bztokzppg}{B^0 \rightarrow K^0\pi^+\pi^-\gamma}

\newcommand{\koneig}{\konei\gamma}
\newcommand{\koneiig}{\koneii\gamma}
\newcommand{\ktwostg}{\ktwost\gamma}

\newcommand{\koneipg}{\koneip\gamma}
\newcommand{\koneiipg}{\koneiip\gamma}
\newcommand{\ktwostpg}{\ktwostp\gamma}

\newcommand{\btokoneg}{B \to K_1 \gamma}
\newcommand{\btokoneig}{B \rightarrow K_1 (1270) \gamma}
\newcommand{\bptokoneipg}{B^+ \rightarrow K_1(1270)^+ \gamma}
\newcommand{\bztokoneizg}{B^0 \rightarrow K_1(1270)^0 \gamma}

\newcommand{\bptokoneiipg}{B^+ \rightarrow K_1 (1400)^+ \gamma}
\newcommand{\bztokoneiizg}{B^0 \rightarrow K_1 (1400)^0 \gamma}
\newcommand{\btoktwostg}{B \to K_2^* (1430)\gamma}
\newcommand{\bptoktwostpg}{B^+\to\ktwostpg}
\newcommand{\bptokpphig}{B^+ \to K^+\phi\gamma}

\newcommand{\mpp}{M_{\pi\pi}}
\newcommand{\mkp}{M_{K\pi}}

\newcommand{\btokstarg}{B \rightarrow K^*\gamma}

\newcommand{\btosgam}{b \rightarrow s\gamma}
\newcommand{\BBbar}{B\overline{B}}

\newcommand{\Dbar}{\overline{D}{}}

\newcommand{\Ebeam}{E^*_{\rm beam}}
\newcommand{\Egamma}{E^*_\gamma}
\newcommand{\Ekpp}{E^*_{K\pi\pi}}

\newcommand{\pcmvec}{\vec{p}^{\;*}}

\newcommand{\mbc}{M_{\rm bc}}
\newcommand{\de}{\Delta E}

\newcommand{\pBvec}{\pcmvec_B}
\newcommand{\pgamvec}{\pcmvec_\gamma}
\newcommand{\pkppvec}{\pcmvec_{K\pi\pi}}
\newcommand{\Br}{{\cal B}}
\newcommand{\calL}{{\cal L}}
\newcommand{\calR}{{\cal R}}

\def\Journal#1#2#3#4{{#1} {\bf #2}, #3 (#4)}

\def\NIMA{Nucl. Instrum. Meth. A}

\def\PLB{Phys. Lett. B}
\def\PRL{Phys. Rev. Lett.}
\def\PRD{Phys. Rev. D}

\def\EPJC{Eur. Phys. J. C}
\def\etal{{\it et al.}}

\newcommand{\EM}[1]{\times10^{-#1}}

\newcommand{\Nkpppg}{318\pm22}     
\newcommand{\Nkzppg}{67\pm10}      

\newcommand{\Ekpppg}{ (8.36\pm 0.64)\%}
\newcommand{\Ekzppg}{ (1.82\pm 0.12)\%}

\newcommand{\Bkpppg}{ (2.50\pm{0.18}\pm{0.22})\EM5 }
\newcommand{\Bkzppg}{ (2.43\pm0.36\pm0.34)\EM5 } 

\newcommand{\Skpppg}{20.4\sigma}
\newcommand{\Skzppg}{10.9\sigma}

\newcommand{\NkoneigI}{   102 \pm 22 } 
\newcommand{\EkoneigI}{   (1.56\pm0.12)\% }
\newcommand{\BkoneigIfull}{  
         (4.28\pm0.94{\rm(stat.)}\pm0.43{\rm(syst.)})\EM5 }
\newcommand{\BkoneigI}{  (4.28\pm0.94\pm0.43)\EM5 }

\newcommand{\SkoneigI}{  9.2\sigma }

\newcommand{\NkoneiigII}{   23 \pm{26} } 
\newcommand{\EkoneiigII}{  (2.68\pm0.20)\% }
\newcommand{\BkoneiigII}{ <1.44\EM5 }
\newcommand{\SkoneiigII}{ 1.3\sigma }

\graphicspath{{ps}}

\begin{document}


\newcommand{\VERSION}{20040818a}

\preprint{\vbox{ \hbox{   }
                 \hbox{BELLE-CONF-0411}
                 \hbox{ICHEP04 11-0656} 
}}


\title{ \quad\\[0.5cm] \boldmath Observation of $\bptokoneipg$ }

\affiliation{Aomori University, Aomori}
\affiliation{Budker Institute of Nuclear Physics, Novosibirsk}
\affiliation{Chiba University, Chiba}
\affiliation{Chonnam National University, Kwangju}
\affiliation{Chuo University, Tokyo}
\affiliation{University of Cincinnati, Cincinnati, Ohio 45221}
\affiliation{University of Frankfurt, Frankfurt}
\affiliation{Gyeongsang National University, Chinju}
\affiliation{University of Hawaii, Honolulu, Hawaii 96822}
\affiliation{High Energy Accelerator Research Organization (KEK), Tsukuba}
\affiliation{Hiroshima Institute of Technology, Hiroshima}
\affiliation{Institute of High Energy Physics, Chinese Academy of Sciences, Beijing}
\affiliation{Institute of High Energy Physics, Vienna}
\affiliation{Institute for Theoretical and Experimental Physics, Moscow}
\affiliation{J. Stefan Institute, Ljubljana}
\affiliation{Kanagawa University, Yokohama}
\affiliation{Korea University, Seoul}
\affiliation{Kyoto University, Kyoto}
\affiliation{Kyungpook National University, Taegu}
\affiliation{Swiss Federal Institute of Technology of Lausanne, EPFL, Lausanne}
\affiliation{University of Ljubljana, Ljubljana}
\affiliation{University of Maribor, Maribor}
\affiliation{University of Melbourne, Victoria}
\affiliation{Nagoya University, Nagoya}
\affiliation{Nara Women's University, Nara}
\affiliation{National Central University, Chung-li}
\affiliation{National Kaohsiung Normal University, Kaohsiung}
\affiliation{National United University, Miao Li}
\affiliation{Department of Physics, National Taiwan University, Taipei}
\affiliation{H. Niewodniczanski Institute of Nuclear Physics, Krakow}
\affiliation{Nihon Dental College, Niigata}
\affiliation{Niigata University, Niigata}
\affiliation{Osaka City University, Osaka}
\affiliation{Osaka University, Osaka}
\affiliation{Panjab University, Chandigarh}
\affiliation{Peking University, Beijing}
\affiliation{Princeton University, Princeton, New Jersey 08545}
\affiliation{RIKEN BNL Research Center, Upton, New York 11973}
\affiliation{Saga University, Saga}
\affiliation{University of Science and Technology of China, Hefei}
\affiliation{Seoul National University, Seoul}
\affiliation{Sungkyunkwan University, Suwon}
\affiliation{University of Sydney, Sydney NSW}
\affiliation{Tata Institute of Fundamental Research, Bombay}
\affiliation{Toho University, Funabashi}
\affiliation{Tohoku Gakuin University, Tagajo}
\affiliation{Tohoku University, Sendai}
\affiliation{Department of Physics, University of Tokyo, Tokyo}
\affiliation{Tokyo Institute of Technology, Tokyo}
\affiliation{Tokyo Metropolitan University, Tokyo}
\affiliation{Tokyo University of Agriculture and Technology, Tokyo}
\affiliation{Toyama National College of Maritime Technology, Toyama}
\affiliation{University of Tsukuba, Tsukuba}
\affiliation{Utkal University, Bhubaneswer}
\affiliation{Virginia Polytechnic Institute and State University, Blacksburg, Virginia 24061}
\affiliation{Yonsei University, Seoul}
  \author{K.~Abe}\affiliation{High Energy Accelerator Research Organization (KEK), Tsukuba} 
  \author{K.~Abe}\affiliation{Tohoku Gakuin University, Tagajo} 
  \author{N.~Abe}\affiliation{Tokyo Institute of Technology, Tokyo} 
  \author{I.~Adachi}\affiliation{High Energy Accelerator Research Organization (KEK), Tsukuba} 
  \author{H.~Aihara}\affiliation{Department of Physics, University of Tokyo, Tokyo} 
  \author{M.~Akatsu}\affiliation{Nagoya University, Nagoya} 
  \author{Y.~Asano}\affiliation{University of Tsukuba, Tsukuba} 
  \author{T.~Aso}\affiliation{Toyama National College of Maritime Technology, Toyama} 
  \author{V.~Aulchenko}\affiliation{Budker Institute of Nuclear Physics, Novosibirsk} 
  \author{T.~Aushev}\affiliation{Institute for Theoretical and Experimental Physics, Moscow} 
% ...
\collaboration{The Belle Collaboration}



\begin{abstract}
We report the observation of the radiative decay $\bptokoneipg$ using a
data sample of $140\invfb$ taken at the $\Upsilon(4S)$ resonance with
the Belle detector at the KEKB $\ee$ collider.  We find the branching
fraction to be $\Br(\bptokoneipg)=\BkoneigIfull$ with a statistical
significance of $\SkoneigI$.  We find no significant signal for
$\bptokoneiipg$ and set an upper limit $\Br(\bptokoneiipg)\BkoneiigII$ at
the 90\% confidence level.  We also measure inclusive branching
fractions for $\bptokpppg$ and $\bztokzppg$ in the mass range
$1\GeVcc<\mkpp<2\GeVcc$.
\end{abstract}


\maketitle

\tighten

{\renewcommand{\thefootnote}{\fnsymbol{footnote}}}
\setcounter{footnote}{0}


Radiative $B$ decays that occur through the flavor changing neutral
current process $\btosgam$ have been one of the most sensitive probes to
search for physics beyond the Standard Model (SM).  The first observed
exclusive radiative decay mode was $\btokstarg$~\cite{kstg,k892}, which
accounts for around 15\% of the total $\btosgam$ branching fraction.
The second mode was $\btoktwostg$, for which evidence was reported by
CLEO and Belle~\cite{kstg,kxgam}.  No other exclusive radiative decay
mode into a two-body final state has been reported; for multi-body final
states, Belle has observed $\bptokpppg$~\cite{CC} which is almost
saturated by $K^{*0}\pi^+\gamma$ and $K^+\rho^0\gamma$ using a
$29\invfb$ data sample~\cite{kxgam}, and $\bptokpphig$ using
$90\invfb$~\cite{kphigam}.

According to theoretical predictions~\cite{theor,theor2}, the branching
fraction of unobserved modes among such exclusive decays should be
within the current experimental sensitivity.  Some of these two-body
decay modes are particularly interesting.  For example, $B \rightarrow
K_1(1270) \gamma$ and $B \rightarrow K_1(1400) \gamma$ ($K_1 \rightarrow
K\pi\pi$) can be used to measure the photon helicity, which may differ
from the SM prediction in some models beyond the SM~\cite{gronau}.  The
neutral decay mode $\bztokoneizg$, $\konei^0\to\ks\rho^0$ would be a
useful channel to measure time-dependent $CP$ violation, which is also
sensitive to the photon helicity in radiative $B$ decay.

In this paper, we report the observation of $\bptokoneipg$, which
is the first radiative $B$ meson decay mode that involves an
axial-vector resonance.  We study radiative decays in the $\kpppg$ and
$\ksppg$ final states, where we search for resonant structure in the
$\kppp$ system.  We also report inclusive measurements of $\bptokpppg$
and $\bztokzppg$, and the results of a search for $\bptokoneiipg$.  The
analysis is based on a data sample of $140\invfb$ taken at the
$\Upsilon(4S)$ resonance with the Belle detector at the KEKB
$e^+e^-$collider~\cite{KEKB}.  The data sample contains 152 million
$\BBbar$ pairs.

The Belle detector is a large-solid-angle magnetic spectrometer that
consists of a three-layer silicon vertex detector (SVD), a 50-layer
central drift chamber (CDC), an array of aerogel threshold \v{C}erenkov
counters (ACC), a barrel-like arrangement of time-of-flight
scintillation counters (TOF), and an electromagnetic calorimeter
comprised of CsI(Tl) crystals (ECL) located inside a super-conducting
solenoid coil that provides a 1.5~T magnetic field.  An iron flux-return
located outside of the coil is instrumented to detect $K_L^0$ mesons and
to identify muons (KLM).  The detector is described in detail
elsewhere~\cite{Belle}.


The photon candidate is the highest energy photon cluster measured with
the barrel ECL ($33^\circ < \theta_{\gamma} < 128^\circ$ in the
laboratory frame) and is required to be an isolated electromagnetic
shower, i.e. 95\% of its energy be concentrated in an array of central
$3\times3$ crystals out of $5\times5$ crystals and the shower width be
less than $5\cm$.  In order to reduce the background due to photons from
$\pi^0$ and $\eta$ decays, we combine the photon candidate with all
other photon clusters in the event with the energy greater than $30\MeV$
($200\MeV$) and reject the event if the invariant mass of any pair is
within $\pm18\MeVcc$ ($\pm32\MeVcc$) around the nominal $\pi^0$ ($\eta$)
mass.  We refer to this requirement as the $\pi^0/\eta$ veto.

Charged tracks, that are reconstructed with the CDC and SVD, are
required to have the momentum in the center-of-mass (c.m.) frame greater
than $200\MeVc$ and to have an impact parameter relative to the
interaction point of less than $5\cm$ along the positron beam axis and
less than $0.5\cm$ in the plane that is transverse to this axis.  These
tracks are identified as pion or kaon candidates by a likelihood ratio
based on the combined information from the ACC and TOF systems and the
$dE/dx$ measurement in the CDC.  We require the kaon likelihood ratio
larger than 0.6 for kaons; the remaining tracks are considered as pion
candidates.  In addition, we remove kaon and pion candidates if they are
identified as an electron, muon or proton.

For neutral kaons, we use $\ks\to\pi^+\pi^-$ candidates that have
invariant masses within $\pm30\MeVcc$ of the $\ks$ mass and have the
c.m.\ momentum greater than $200\MeVc$.
We do not apply the particle identification criteria for these pions.
The two pions are required to have a common vertex that is displaced
from the interaction point. The $\ks$ momentum direction is also
required to be consistent with the $\ks$ flight direction.

We select $\kppp$ and $\kspp$ combinations in the mass range
$1\GeVcc<\mkpp<2\GeVcc$.  We combine this $\kpp$ system and the photon
candidate, and identify $B$ meson candidates using the following two
independent kinematic variables: the beam-energy constrained mass
$\mbc\equiv \sqrt{(\Ebeam/c^2)^2 - |\pBvec/c|^2 }$ and the energy
difference $\de\equiv \Ekpp + \Egamma - \Ebeam$, where $\Ebeam$ is the
beam energy and $\pBvec$ is the momentum of the $B$ candidate in the
c.m.\ frame. (Variables with an asterisk are calculated in the
c.m.\  frame.)  The momentum $\pBvec$ is calculated without using the
absolute value of the photon momentum according to
\begin{equation}
\pBvec = \pkppvec + \frac{\pgamvec}{\Egamma} \times (\Ebeam - \Ekpp)
\end{equation}
since the $\kpp$ momentum $\pkppvec$ and the beam energy are determined
with substantially better precision than that of the primary photon.

We select $B$ candidates within $-0.1\GeV<\de<0.08\GeV$ and
$\mbc>5.2\GeVcc$.  If there exist multiple candidates, we choose the
candidate that has the highest confidence level for the $\kppp$
($\pi^+\pi^-$ for the $\kspp$ mode) vertex fit.


The continuum background ($\ee\to q\bar{q}$, $q=u,d,s,c$) is the
dominant background in this analysis.  To separate signal events from
continuum background events, we use two variables: the $B$ flight
direction ($\cos\theta_B$) and a Fisher discriminant~\cite{fisher}
constructed from a set of shape variables~\cite{KSFW}.  We determine the
probability density functions for signal and continuum background for
each of these variables, and combine them into a likelihood ratio.

The $\cos\theta_B$ distribution of the signal events follows a
$1-\cos^2\theta_B$ distribution while that of $q\bar{q}$ has a nearly
uniform distribution.  The likelihood function for the flight direction
$\calL_{S(B)}^{\cos\theta_B}$ is modeled as a second (first) order
polynomial of $\cos\theta_B$ for the signal (continuum background).

For the shape variables, we use 16 modified Fox-Wolfram
moments~\cite{KSFW,fox-wolfram} that are calculated from the particle
momenta in the following four categories: 1) particles that form the
signal candidate, 2) the remaining charged particles, 3) the remaining
neutral particles, and 4) a hypothetical particle for the missing
momentum of the event.  The Fisher discriminant is calculated from these
moments and the scalar sum of the transverse momentum.  The likelihood
function for this Fisher discriminant $\calL_{S(B)}^{\rm Fisher}$ is
modeled by a bifurcated Gaussian function both for the signal and for
the continuum background.

These likelihood functions are then combined using a single likelihood
ratio,
\begin{equation}
\calR_S = \frac{\calL_S^{\cos\theta_B} \calL_S^{\rm Fisher}}{
                \calL_S^{\cos\theta_B} \calL_S^{\rm Fisher} +
                \calL_B^{\cos\theta_B} \calL_B^{\rm Fisher}}\,.
\end{equation}
We determine the $\calR_S$ requirement by maximizing
$N_S\left/\sqrt{N_S+N_B}\right.$, where $N_S$ and $N_B$ are the expected
number of the signal and background events, respectively.  For this
purpose, we use four signal Monte Carlo (MC) samples, $\bptokoneipg$ and
$\bztokoneizg$ in the
$\kpp$ mass range $1.1\GeVcc<\mkpp<1.5\GeVcc$, and $\bptokoneiipg$ and
$\bztokoneiizg$ in
$1.225\GeVcc<\mkpp<1.575\GeVcc$, assuming all the $\btokoneg$ branching
fractions are $1\EM5$.  We find a requirement of $\calR_S>0.9$ is
optimal for all four cases.


The signal yields for $\bptokpppg$ and $\bztokzppg$ are extracted from a
binned maximum likelihood fit to the $\mbc$ distribution with signal and
background components.  In addition to the continuum background, we
consider four $B$ decay background sources: a collection of known $B$
decays through the $b\to c$ transition (referred to as the $b\to c$
background), hadronic $B$ decays through the $b\to u$, $b\to s$ and
$b\to d$ transitions (charmless background), $\btokstarg$ background,
and radiative decays to the final states other than $K^*\gamma$ and
$\kppg$ (other $\btosgam$ background).  In order to suppress the
$\btokstarg$ background, we reject the event if $\de$ and $\mbc$
calculated from the $K\pi\gamma$ combination is within
$-0.2\GeV<\de<0.1\GeV$ and $\mbc>5.27\GeVcc$.

The signal $\mbc$ distribution is modeled with a single Gaussian
function at the $B$ meson mass with a resolution of $2.6\MeVcc$, that is
calibrated using a $B^0\to D^-\pi^+$, $D^-\to K^+\pi^-\pi^-$ ($B^+\to
\Dbar^0\pi^+$, $\Dbar^0\to \ks\pi^+\pi^-$) sample for the $\kpppg$
($\ksppg$) final state.  We model the background $\mbc$ distributions
using large samples of MC events.  We find the sum of the continuum and
$b\to c$ backgrounds is described by an ARGUS function~\cite{ARGUS}.
The ARGUS shape parameter is taken from the MC sample, which is
consistent with the $\de$ sideband data in $0.1\GeV<\de<0.2\GeV$ where
$B$ decay backgrounds are negligible.  Charmless decays, $B\to
K^*\gamma$ and other $\btosgam$ decays each produce a small signal-like
component on top of the background-like component.  We model each of
these backgrounds as a sum of an ARGUS function and a Gaussian function.
The size of the continuum plus $b\to c$ background is floated in the
fit; the size of the other three components are fixed in the fit.

\begin{figure}[ht]
\includegraphics[width=0.9\textwidth]{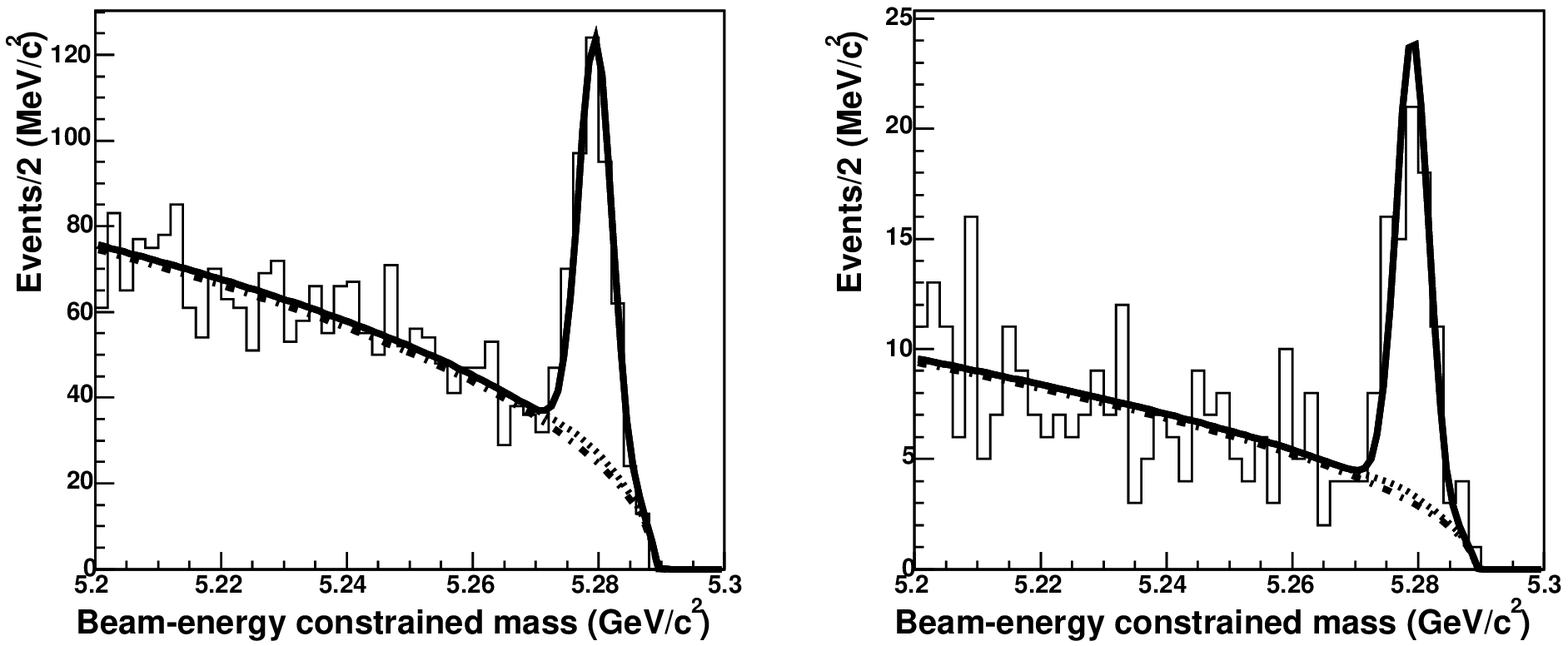}
\caption{$\mbc$ distributions for $\bptokpppg$ (left) and $\bztokzppg$
         (right). Lines show the continuum plus $b\to c$ background
         component (dot-dashed), total background (dotted) and the total
         fit result (solid). }
\label{mbc}
\end{figure}

The fit result is shown in Fig.~\ref{mbc}.  For the $\bptokpppg$ mode, we
obtain $\Nkpppg$ events with a statistical significance of $\Skpppg$,
where the significance is defined as $\sqrt{-2\ln({\calL}_0/{\calL}_{\rm
max})}$, and $\calL_{\rm max}$ and $\calL_0$ denote the maximum
likelihoods of the fit with and without the signal component,
respectively.  Similarly, we obtain $\Nkzppg$ events with a statistical
significance of $\Skzppg$ for the $\bztokzppg$ mode.

We evaluate the following systematic errors in the fitting procedure.
The width and the peak position of the signal Gaussian are varied by
$\pm1\sigma$, where $\sigma$ is the error from the $B\to D\pi$
calibration sample.  For the continuum plus $b\to c$ background, we vary
the ARGUS shape parameter by $\pm1\sigma$, where $\sigma$ is determined
from the $\de$ sideband data.  The $\btokstarg$ component is varied by
the branching fraction uncertainty.  The other $\btosgam$ component is
varied by the uncertainty in the branching fraction \cite{HFAG} and the
fraction to the $\kppg$ final state \cite{acpxsgam}.  For the charmless
background, we vary the normalization by $\pm100\%$.  We take the sum in
quadrature of all the deviations in the signal yield of these tests.
The results are given in Table~\ref{table:sys1}.

\begin{table}[bh]
\caption{Fitting systematic errors for $\bptokpppg$ and $\bztokzppg$.}
\label{table:sys1}
\begin{tabular}{l@{~~~}c@{~~~}c}
\hline 
{}  & $\bptokpppg$ & $\bztokzppg$ \\
\hline 
Signal $\mbc$ parameters  & $(+0.65/{-}0.68)\%$   & $(+1.8/{-}2.2)\%$ \\
Background ARGUS shape    & $(+2.4/{-}2.6)\%$   & $(+4.0/{-}4.5)\%$ \\
$\btokstarg$              & $(+0.25/{-}0.11)\%$ & $(+0.01/{-}0.17)\%$ \\
Other $\btosgam$          & $(+0.36/{-}0.54)\%$   & $(+0.84/{-}0.96)\%$ \\
Charmless background      & $(+0.60/{-}0.60)\%$ & $(+0.74/{-}0.74)\%$ \\
\hline 
MC stat. error            & $(+4.8/{-}4.8)\%$   & $(+11.5/{-}11.0)\%$ \\
\hline 
Total fitting error       & $(+5.5/{-}5.6)\%$ & $(+12.4/{-}12.1)\%$ \\
\hline 
\end{tabular}
\end{table}


In order to decompose intermediate resonances that may be involved in
the $\kppp$ final state, we perform an unbinned maximum likelihood fit to
the $\mbc$ and $\mkpp$ distributions of the $\bptokpppg$ candidates.
We do not use $\bztokzppg$ due to the limited statistics.  There are
many possible resonances that can contribute: $\konei$, $\koneii$,
$\ktwost$, $K^*(1410)$, $K^*(1680)$, and so on.  Here we consider the
first three resonances, and include an additional non-resonant
$\bptokpppg$ component whose $\mkpp$ distribution is modeled with a set
of two broad Gaussian functions, which are determined from an inclusive
$B\to X_s\gamma$ MC sample.  The $\btoktwostg$ component, which is
already measured, is fixed in the fit.  We model the $\konei$ resonance
as a sum of three decay chains, $\konei^+\to K^+\rho^0$,
$\rho^0\to\pi^+\pi^-$; $\konei^+\to K^{*0}\pi^+$, $K^{*0}\to K^+\pi^-$;
and $\konei^+\to K_0^*(1430)\pi^+$, $K_0^*(1430)\to K^+\pi^-$.  We do
not consider interference between these final states.  The $\mkpp$
distribution for each decay chain is described by convolving the two
relativistic Breit-Wigner functions of the decay chain.  The $\koneii$
resonance is modeled with a single decay chain, $\koneii^+\to
K^{*0}\pi^+$, $K^{*0}\to K^+\pi^-$.  The $\mkpp$ distributions of other
$\btosgam$ and continuum plus $b\to c$ backgrounds are modeled by using
a function $(p_0 + p_1 x) e^{p_2 + p_3 x + p_4 x^2}$, where $x=\mkpp$,
and $p_i$ $(i=0\ldots4)$ are empirical parameters that are determined by
using MC samples.

\begin{figure}[ht]
\includegraphics[width=0.65\textwidth]{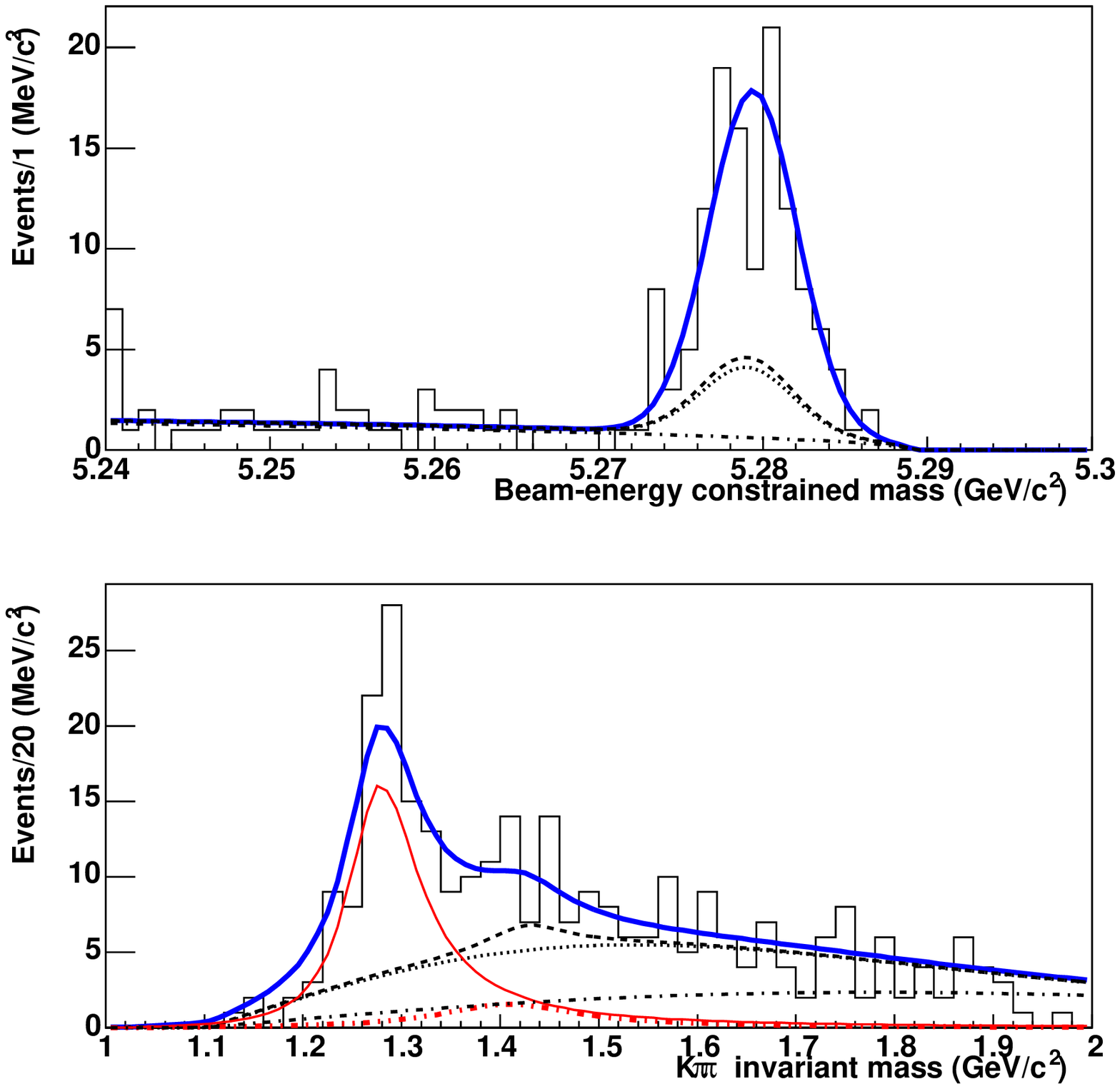}
\caption{$\mbc$ distribution for $1.2\GeVcc<\mkpp<1.4\GeVcc$ (top) and
          $\mkpp$ distribution for $\mbc>5.27\GeVcc$ (bottom) of the
          $\koneig$ enriched sample with $0.6\GeVcc<\mpp<0.9\GeVcc$.
          Lines show the projections of the fit results for the
          continuum plus $b\to c$ background component (dot-dashed),
          total background without and with the $\ktwostg$ component
          (dotted and dashed, respectively), $\koneig$ (thin solid line)
          and $\koneiig$ (dot-dot-dashed) components, and the sum of
          all components (thick solid line). }
\label{fig:k1270g}
\end{figure}

In order to enhance the $\konei$ component, we further select the events
with the $\pi^+\pi^-$ mass in the $\rho^0$ mass region,
$0.6\GeVcc<\mpp<0.9\GeVcc$ (the $\konei\to K\rho$ branching fraction
$(42\pm6)\%$ being much larger than the $\koneii\to K\rho$ branching
fraction $(3\pm3)\%$).  Even with this requirement, substantial
$K^*\pi\gamma$ and $K_0^*(1430)\pi\gamma$ events remain in the sample
due to the overlapping kinematics of the final states.  The fit result
is shown in Fig.~\ref{fig:k1270g}.  We find $\NkoneigI$ events for the
$\bptokoneipg$ component in the mass range $1\GeVcc<\mkpp<2\GeVcc$ with
a statistical significance of $\SkoneigI$.  We find the $\koneiig$
contribution is much smaller than that from $\koneig$, and is consistent
with zero.

\begin{figure}[ht]
\includegraphics[width=0.65\textwidth]{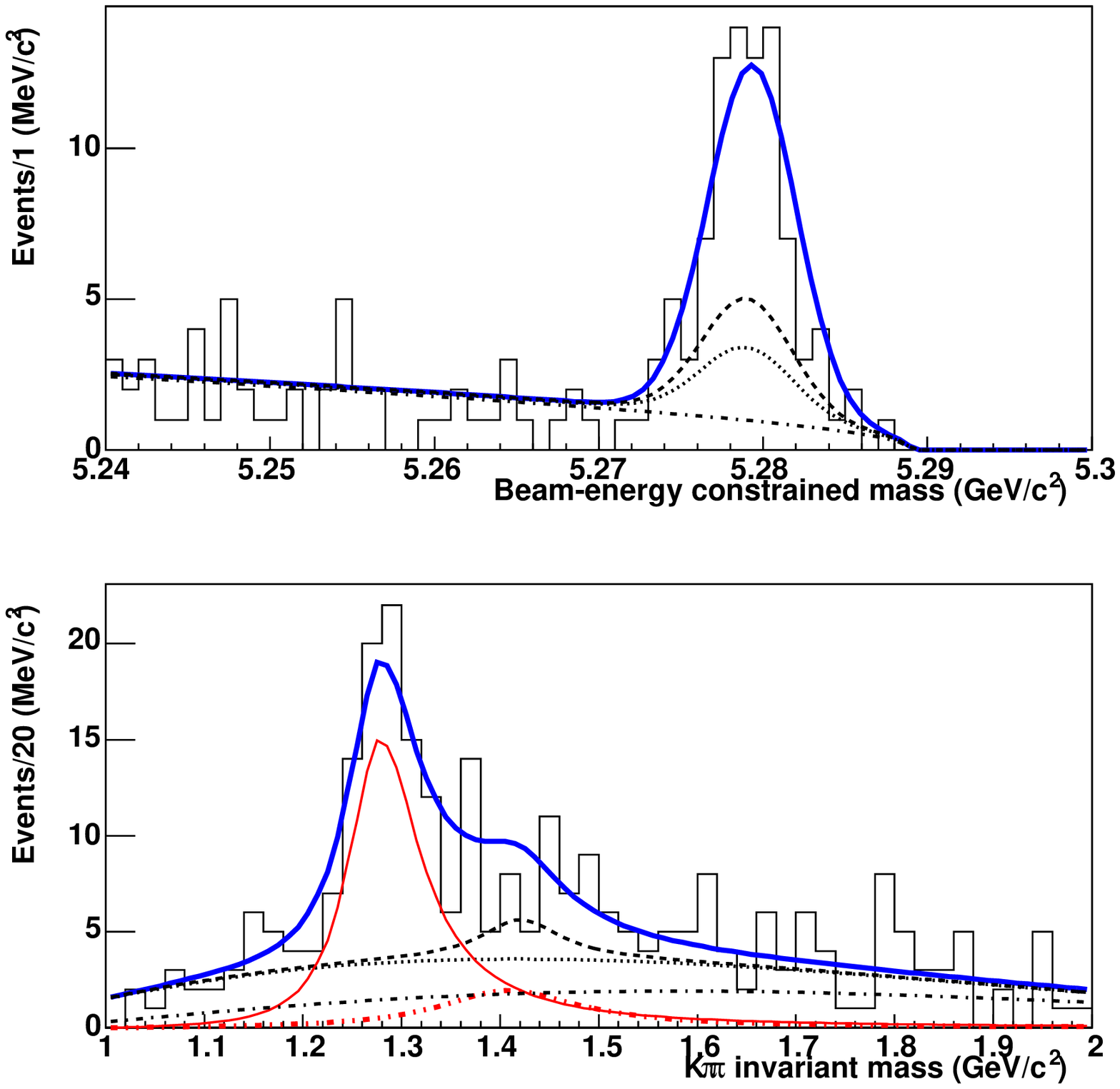}
\caption{$\mbc$ distribution for $1.2\GeVcc<\mkpp<1.4\GeVcc$ (top) and
          $\mkpp$ distribution for $\mbc>5.27\GeVcc$ (bottom) of the
          $\koneiig$ enriched sample with $0.8\GeVcc<\mkp<1.0\GeVcc$.
          Lines show the projections of the fit results with the same
          definition as in Fig.~\ref{fig:k1270g}. }
\label{fig:k1400g}
\end{figure}

Similarly, we select the events with the $K^+\pi^-$ mass in the $K^{*0}$
mass region, $0.8\GeVcc<\mkp<1.0\GeVcc$, in order to enhance the
$\koneii$ component (the $\koneii\to K^*\pi$ branching fraction
$(94\pm6)\%$ being much larger than the $\konei\to K^*\pi$ branching
fraction $(16\pm5)\%$).  The fit result is shown in
Fig.~\ref{fig:k1400g}.  We still find a sizable $\koneig$ component,
and only a small $\koneiig$ component.  We find $\NkoneiigII$ events for
the $\bptokoneiipg$ component in the mass range $1\GeVcc<\mkpp<2\GeVcc$
with a statistical significance of $\SkoneiigII$.  Since the $\koneiig$
component is not significant, we set a 90\% confidence level upper limit
on the signal yield, $N_{90}$, which is calculated from the relation
$\int_0^{N_{90}} \calL(n)dn=0.9\int_0^\infty \calL(n)dn$, where
$\calL(n)$ is the likelihood function with the signal yield fixed at
$n$.

Additional tests are performed to evaluate the fitting systematic
errors, in addition to those previously mentioned.  We vary the $\konei$
and $\koneii$ signal peak positions and widths according to the errors
given by the Particle Data Group (PDG)~\cite{PDG}.  We vary the
$\ktwostg$ branching fraction according to the PDG error.  These fitting
systematic errors are summarized in Table~\ref{table:sys2}.

\begin{table}[bh]
\caption{Fitting systematic errors for $\bptokoneipg$ and $\bptokoneiipg$.}
\label{table:sys2}
\begin{tabular}{l@{~~~}c@{~~~}c}
\hline 
{} & $\bptokoneipg$ & $\bptokoneiipg$ \\
\hline 
Signal $\mbc$ parameters  & $(+0.35/{-}0.14)\%$   & $(+0.46/{-}0.50)\%$ \\
Background ARGUS shape    & $(+0.33/{-}0.36)\%$   & $(+0.31/{-}0.32)\%$ \\
$\btokstarg$              & $(+0.01/{-}0.02)\%$ & $(+0.63/{-}0.27)\%$ \\
Other $\btosgam$          & $(+0.03/{-}0.00)\%$   & $(+0.05/{-}0.07)\%$ \\
Charmless background      & $(+0.01/{-}0.02)\%$ & $(+0.13/{-}0.13)\%$ \\
\hline 
Breit-Wigner line shape   & $(+0.95/{-}1.0)\%$ & $(+4.4/{-}5.1)\%$ \\
$\Br(\bptoktwostpg)$      & $(+0.60/{-}0.70)\%$ & $(+28.6/{-}28.6)\%$ \\
\hline 
MC stat. error            & $(+6.6/{-}6.4)\%$  & $(5.3/{-}5.2)\%$ \\
\hline 
Total fitting error       & $(+6.7/{-}6.5)\%$ & $(+29.4/{-}29.5)\%$ \\
\hline 
\end{tabular}
\end{table}


The reconstruction efficiency is obtained from the signal MC samples.
For the inclusive measurement, we consider the efficiency difference
between the different intermediate state assumptions; we use a weighted
average of the $\bptokpppg$ efficiencies for non-resonant decays and the
one through the $\koneip$ state, using the measured ratio of the
$\koneipg$ and non-resonant $\kpppg$ events ($\koneiipg$ and $\ktwostpg$
events are neglected).  We assume the same ratio for the $\bztokzppg$
efficiency.  We estimate the systematic errors due to photon detection
(2.8\%), tracking ($\sim1\%$ per track), charged particle identification
(1--2\% per particle), and $\ks$ reconstruction (4.5\%) from control
samples of radiative Bhabha events, partially reconstructed $D^{*+}\to
D^0\pi^+$, $D^0\to\ks\pi^+\pi^-$ events, $D^{*+}$-tagged $D^0\to
K^-\pi^+$ decays, and the ratio of $D^+\to\ks\pi^+$ decays to $D^+\to
K^-\pi^+\pi^+$ decays, respectively.  The systematic error due to the
likelihood ratio and $\pi^0/\eta$ veto requirements is estimated to be
6.1\% (3.2\%) from a control sample of $B^0\to D^+\pi^-$, $D^+\to
K^-\pi^+\pi^+$ ($B^+\to \Dbar^0\pi^+$, $\Dbar^0\to\ks\pi^+\pi^-$) events
for the $\bptokpppg$ ($\bztokzppg$) mode.  The efficiencies and their
systematic errors are summarized in Table~\ref{table:eff}.

\begin{table}[bh]
\caption{Relative systematic errors and total efficiencies.}
\label{table:eff}
\begin{tabular}{lcccc}
\hline 
{} & \multicolumn{3}{c}{$\bptokpppg$} & $\bztokzppg$ \\
{} & $\koneipg$ & $\koneiipg$ & inclusive & inclusive \\
\hline
Photon detection & $\pm2.8\%$ & $\pm2.8\%$ & $\pm2.8\%$ & $\pm2.8\%$ \\
Tracking         & $\pm3.0\%$ & $\pm3.0\%$ & $\pm3.0\%$ & $\pm2.1\%$\\
Particle id.     & $\pm2.1\%$ & $\pm2.1\%$ & $\pm2.1\%$ & $\pm1.0\%$\\
$\ks$ reconstruction   & --- & --- & --- & $\pm2.4\%$\\
$\calR_S$ and $\pi^0/\eta$ veto
                 & $\pm6.1\%$ & $\pm6.1\%$ & $\pm6.1\%$ & $\pm3.2\%$\\
\hline 
Total efficiency error
                 & $\pm 7.6\%$ & $\pm 7.6\%$ & $\pm 7.6\%$ & $\pm 6.6\%$\\
\hline 
Efficiency & ~$\EkoneigI$~ & ~$\EkoneiigII$~ & ~$\Ekpppg$~ & ~$\Ekzppg$~ \\
\hline 
\end{tabular}
\end{table}


Using the signal yield and reconstruction efficiency for $\bptokoneipg$,
and the number of $B$ meson pairs ($(152.0\pm0.6)\times10^6$), we find
\begin{equation}
\Br(\bptokoneipg) = \BkoneigIfull.
\end{equation}
Similarly, we measure the inclusive branching fractions of $\bptokpppg$
and $\bztokzppg$ in the mass range $1\GeVcc<\mkpp<2\GeVcc$,
\begin{equation}
\begin{array}{r@{~$=$~}l}
\Br(\bptokpppg)&\Bkpppg,\\
\Br(\bztokzppg)&\Bkzppg.
\end{array}
\end{equation}
For $\bptokoneiipg$, we use the upper limit on the signal yield to which
the fitting systematic error is added, and the systematic error
subtracted efficiency.  We obtain an upper limit,
\begin{equation}
\Br(\bptokoneiipg) \BkoneiigII \mbox{~~(90\% C.L.)}.
\end{equation}
These results are summarized in Table~\ref{table:results}.

\begin{table}[th]
\caption{Results on the signal yields, efficiencies, branching fractions
and significances.}
\label{table:results}
\begin{tabular}{lcccc}
\hline 
 & yield & efficiency & branching fraction & significance \\
\hline 
$\bptokoneipg$  & $\NkoneigI$ & $\EkoneigI$ & $\BkoneigI$ & $\SkoneigI$ \\
$\bptokoneiipg$ &
            $\NkoneiigII$ & $\EkoneiigII$ & $\BkoneiigII$ & $\SkoneiigII$ \\
$\bptokpppg$ & ~$\Nkpppg$~ & ~$\Ekpppg$~ & ~$\Bkpppg$~ & $\Skpppg$ \\
$\bztokzppg$ & ~$\Nkzppg$~ & ~$\Ekzppg$~ & ~$\Bkzppg$~ & $\Skzppg$ \\
\hline 
\end{tabular}
\end{table}


We find the branching fraction for $\bptokoneipg$ is larger than that
for $\bptokoneiipg$.  This difference may be explained by the
$\konei$--$\koneii$ mixing angle~\cite{theor2}, but the branching
fraction for $\bptokoneipg$ itself is much larger than theory
predictions~\cite{theor,theor2}.  Both results are consistent with
previous upper limits~\cite{kxgam}; the $\bptokoneiipg$ upper limit is
significantly reduced.


The inclusive branching fraction $\bptokpppg$ is consistent with the
previous measurement with a significantly improved error~\cite{kxgam}.
The neutral decay mode $\bztokzppg$ is measured to have a similar
branching fraction, which suggests the yet to be measured
$\bztokoneizg$ branching fraction could also be large.


To conclude, we observe a new radiative decay mode, $\btokoneig$, with a
branching fraction of $\BkoneigIfull$.  Although the corresponding
neutral decay is not measured yet, we expect a similar branching
fraction since we have also measured the inclusive branching fractions
for $\bptokpppg$ and $\bztokzppg$ and found that they are close each
other.  With a larger data sample, these decay modes can be used to
search for physics beyond the SM.


We would like to thank the KEKB group for the excellent operation of the
accelerator, the KEK Cryogenics group for the efficient
operation of the solenoid, and the KEK computer group and
the National Institute of Informatics for valuable computing
and Super-SINET network support. We acknowledge support from
the Ministry of Education, Culture, Sports, Science, and
Technology of Japan and the Japan Society for the Promotion
of Science; the Australian Research Council and the
Australian Department of Education, Science and Training;
the National Science Foundation of China under contract
No.~10175071; the Department of Science and Technology of
India; the BK21 program of the Ministry of Education of
Korea and the CHEP SRC program of the Korea Science and
Engineering Foundation; the Polish State Committee for
Scientific Research under contract No.~2P03B 01324; the
Ministry of Science and Technology of the Russian
Federation; the Ministry of Education, Science and Sport of
the Republic of Slovenia; the National Science Council and
the Ministry of Education of Taiwan; and the U.S.\
Department of Energy.

		      
%


\begin{thebibliography}{99}
\bibitem{k892}
%
We use $K^*$ to denote $K^*(892)$ throughout the paper.
\bibitem{CC}
%
Throughout this paper, the inclusion of the charge conjugate mode decay
is implied unless otherwise stated.
\end{thebibliography}
\end{document}